\documentclass[12pt,eadjoint tfnpsf]{article}

\catcode`\@=11
\@addtoreset{equation}{section}

\global\arraycolsep=1pt

\usepackage{amsbsy,amssymb,latexsym,amsfonts, amsmath}
\usepackage{mathrsfs}
\usepackage{graphicx}

\RequirePackage[dvips,usenames]{color}
\definecolor{fireblick}{rgb}{0.698039,0.133333,0.133333}

\newcommand{\beq}{\begin{equation}}
\newcommand{\eeq}{\end{equation}}
\newcommand{\bea}{\begin{eqnarray}}
\newcommand{\eea}{\end{eqnarray}}

\newcommand{\CF}{{\mathcal F}}

\newcommand{\CN}{{\mathcal N}}
\newcommand{\CO}{{\mathcal O}}
\newcommand{\CP}{{\mathcal P}}

\newcommand\C{{\cal C}}

\def\Tr{\mathop{\rm Tr}}
\newcommand\tr{\mathrm{tr}}

\renewcommand{\thefootnote}{\fnsymbol{footnote}}

\begin{document}

%
%
\begin{titlepage}

\begin{flushright}
\normalsize
~~~~
SISSA  18/2011/EP-FM\\
\end{flushright}

\vspace{80pt}

\begin{center}
{\LARGE Quantum Hitchin Systems via $\beta$-deformed Matrix Models}\\
\end{center}

\vspace{25pt}

\begin{center}
{
Giulio Bonelli, Kazunobu Maruyoshi and Alessandro Tanzini
}\\
%
\vspace{15pt}
%
International School of Advanced Studies (SISSA) \\via Bonomea 265, 34136 Trieste, Italy 
and INFN, Sezione di Trieste \\
\end{center}
%
\vspace{20pt}
\begin{center}
Abstract\\
\end{center}
We study the quantization of Hitchin systems in terms of $\beta$-deformations of generalized matrix models
related to conformal blocks of Liouville theory on punctured Riemann surfaces.
We show that in a suitable limit, corresponding to the Nekrasov-Shatashvili one, 
the loop equations of the matrix model reproduce the Hamiltonians of 
the {\it quantum} Hitchin system on the sphere and the torus with marked points. 
The eigenvalues of these Hamiltonians are shown to be the $\epsilon_1$-deformation of the 
chiral observables of the corresponding ${\cal N}=2$ four dimensional gauge theory. 
Moreover, we find the exact wave-functions 
in terms of the matrix model representation of the conformal blocks with degenerate field insertions.

\vfill

\setcounter{footnote}{0}
\renewcommand{\thefootnote}{\arabic{footnote}}

\end{titlepage}

\section{Introduction}
\label{sec:intro}
  
${\cal N}=2$ supersymmetric gauge theories in four dimensions display very interesting mathematical structures
in their supersymmetrically saturated sectors. These structures allow an exact characterization
of several important physical aspects, such as their low energy behavior and stable spectra.
These data are encoded in the celebrated Seiberg-Witten solution \cite{SW}. It was soon realized that
the Seiberg-Witten data can be recovered from integrable systems in terms of their spectral curves \cite{Martinec}.
In this context the Hitchin integrable system has emerged as the fundamental geometric structure underlying 
the M-theory description of ${\cal N}=2$ theories \cite{DW,G,GMN}. 

On the other hand the Seiberg-Witten solution can also be recovered, at least in the case of linear and elliptic quiver 
${\cal N}=2$ theories,
via equivariant localization on the instanton moduli space \cite{N,P}.
Indeed, this approach contains further information encoded in the expansion in the equivariant parameters of the 
$\Omega$-background.
This opens the issue of relating the full Nekrasov partition function to a suitable quantization of the Hitchin system.

  A crucial result in this context is provided by the AGT correspondence \cite{AGT} 
  relating the Nekrasov partition function to conformal blocks of Liouville/Toda field theories in two dimensions.
  In \cite{cicca,Teschner:2010je} it was proposed that this correspondence should be regarded 
  as a two parameter quantization of the Hitchin system itself, or, in field theory language, as its second quantization. 
  Here we will address these issues in the particular limiting case 
  in which one of the two equivariant parameters is vanishing. 
  This was identified by Nekrasov and Shatashvili \cite{NS} to provide the first quantization of the integrable 
  system\footnote{See also \cite{MM, MM2, Popolitov, Nekrasov:2010ka, Orlando:2010uu, He:2010xa, Kozlowski:2010tv, 
                  Poghossian:2010pn, MMM,Yamada, Piatek:2011tp, Nekrasov:2011bc, Fucito:2011pn, Zenkevich:2011zx, 
                  Dorey:2011pa, Chen:2011sj} 
  for the relation between the gauge theory and the quantized integrable system.}.
  In the context of AGT correspondence the instanton partition function 
  in the Nekrasov-Shatashvili limit can be related to the insertion of degenerate fields 
  in the Liouville theory \cite{AT,MT}, which corresponds to the insertion of surface operators 
  in the gauge theory side \cite{AGGTV, DGH, sara, taki, manabe, Kozcaz, zhao}.
  
  In our approach, we will make use of the matrix model perspective on AGT correspondence developed in \cite{DV}. 
  This was further elaborated for Liouville theory on the sphere 
  in \cite{IMO, EM1, mm, MMS1, Fujita:2009gf, Sulkowski, MMS2, IO, MMMorozov, EM2, Itoyama:2010na, Brini:2010fc, CDV, 
           Santillan:2011fp, Mironov:2011jn, Itoyama:2011mr}, 
  on the torus in \cite{MY, MMS} and in \cite{BMTY} at all genera.
  In this context the equivariant parameters of the Nekrasov partition function are encoded in the $\beta$-deformation 
  of the standard Van-der-Monde measure \cite{DV}.
  Notice that for $\beta$-deformed matrix models the algebraic equation defining the spectral curve
  gets deformed into a {\it differential} equation 
  which can be interpreted as a Schr\"odinger equation \cite{EO, Chekhov:2009mm, CEM}.
  Our proposal identifies this differential equation, in the Nekrasov-Shatashvili limit, 
  as providing the quantum Hamiltonians of the associated Hitchin integrable system.
  Moreover the associated wave-function is described in terms of the $\beta$-deformed generalized matrix model 
  corresponding to degenerate field insertions in the Liouville theory \cite{AGGTV, DGH, MT}.

  Let us notice also that the quantization of Hitchin systems plays a vital r\^ole in the context of Langlands duality
  \cite{frenkel}.

  The organization of this paper is as follows.
  In section \ref{sec:Hitchin}, we review the M-theory perspective on the Hitchin system 
  and some basic facts on its quantization at low genera, namely the sphere and the torus with marked points.
  We derive the loop equation for the generalized $\beta$-deformed matrix model 
  both in the sphere and torus case in sections \ref{sec:sphere} and \ref{sec:torus} respectively 
  and we show that in the Nekrasov-Shatashvili limit these reproduces the Hamiltonians of the quantized Hitchin system. 
  Moreover, we provide a description of the associated wave-functions 
  in terms of the $\beta$-deformed generalized matrix model describing degenerate field insertions 
  in the Liouville theory \cite{AGGTV, DGH, MT}.
  In section \ref{sec:conclusion} we present some final comments and further directions.

\section{Quantum Hitchin systems {\it in nuce}}
\label{sec:Hitchin}
  In the M-theory framework, the Hitchin system arises by considering the geometry of a system of $N$ M5-branes 
  wrapped on a manifold $Y_6$ with topology $\C\times{\bf R}^4\times \{pt\}$ 
  in $T^*\C\times{\bf R}^4\times{\bf R}^3$ where $\C$ is a Riemann surface. 
  This \cite{DV} should be equipped with a non trivial fibration of ${\bf R}^4$ over $\C$ 
  which specifies the $\Omega$-background of Nekrasov \cite{N}. 
  The geometry of the M5-branes bound state is described by 
  an $N$-fold branched covering of $\C$ given by the algebraic equation
    \beq
    x^N
     =     \sum_{j=2}^N \phi_j(z) x^{N-j}
           \label{1}
   \eeq
  where $x$ is a section of $T^*\C$ and $\phi_j$ are $(j,0)$-holomorphic differentials on $\C$
  whose singularity structure at the punctures identifies the matter content of the gauge theory.
  The quantum mechanics of this latter structure passes by interpreting (\ref{1}) as the spectral curve of the associated 
  classical system encoding the Seiberg-Witten solution (\ref{1}), and then quantizing via a suitable deformation.
  
  In what follows, we focus on the case with two M5-branes and the corresponding Hitchin systems.
  In this case, the M-theory curve is specified by the quadratic differential: $x^2 = W_2(z)$.
  We further focus on the situation where the singularities of the quadratic differential are all of {\it regular} type
  meaning the poles of at most degree $2$.
  There is one kind of regular punctures in this two M5-branes case, so
  the Riemann surface $\C$ is just specified by genus and the number of punctures.
  Thus, we denote this by $\mathcal{C}_{g,n}$.
  Under the particular marking of $\mathcal{C}_{g,n}$, 
  the worldvolume low energy theory is the weakly coupled $SU(2)^{n+ 3g - 3}$ superconformal quiver gauge theory 
  \cite{Witten, G}.
  The Hitchin system is associated with this gauge theory.

In order to have explicit parametrization of the Hamiltonians, 
let's review and specify Hitchin systems and their quantization at low genera, that is sphere and torus with 
an arbitrary number of punctures, following the approach of \cite{gaw}.

Let us start from some general features of the system.
The Higgs field satisfies the condition
$$
\bar\partial_A\Phi=\sum_k \lambda_k\delta_{w_k}
$$
which can be simplified by $\bar A=h^{-1}\bar\partial h$ and $\Phi=h^{-1}\tilde\Phi h$
to 
\beq
\bar\partial\tilde\Phi=\sum_k \nu_k\delta_{w_k},
\label{inter}
\eeq
where $\nu_k=h^{-1}(w_k)\lambda_k h(w_k)$. Eq.(\ref{inter})
admits a unique solution iff $\sum_k\nu_k=0$, given by
\beq
\tilde\Phi=\sum_k \nu_k\, \omega_{w_k,z^*}+\Phi^0
\label{general}
\eeq
where $\omega_{PQ}$ is the unique normalized abelian differential of third kind, i.e.
holomorphic on $\Sigma\setminus\{P,Q\}$ with simple poles at $P$ and $Q$
with residues respectively $+1$ and $-1$ and vanishing A-periods, while
$\Phi^0=\phi^0_I\omega_I$ is a Lie algebra valued holomorphic differential and $\omega_I$ is a basis of normalized 
holomorphic differentials on $\Sigma$.

As explained in \cite{gaw}, the Poisson brackets are induced by the Lie algebra of the complexified gauge group.
At every puncture the residues are expanded as $\nu_k=\sum_a \nu_k^a t^a$, where $t^a$ is a basis of the Lie algebra and
the Poisson brackets are $\{\nu_k^a,\nu_l^b\}_{PB}=i\delta_{kl}f^{abc}\nu_k^c$.

The Hamiltonians are the Chern polynomials of the Higgs field, namely the coefficients of the expansion of the spectral curve (\ref{1})
as
$$
det\left(\Phi-x\cdot {\bf 1}\right)=0.
$$

  The quantization of the integrable system, as proposed in \cite{gaw}, is induced 
  by the quantization of the Poisson brackets above.

  The case of our interest is a projection to the Cartan degrees of freedom of the general integrable system 
  specified to $SL(2,{\bf C})$.

  In the sphere case, there are no holomorphic differentials and
  the Higgs field reads (see also \cite{Teschner:2010je})
$$
\tilde\Phi=\sum_k\frac{\nu_k}{z-w_k}\frac{dz}{2\pi i}
$$
where $\nu_k$ is the only Cartan element.
The corresponding relevant Hamiltonians are generated by
\beq
\Tr\Phi^2= \sum_k \left(\frac{{\bf J}^2_k}{(z-w_k)^2}  + \frac{{\bf H}^{(0)}_k}{z-w_k} \right)
\left(\frac{dz}{2\pi i}\right)^2
\label{HamiltonianC0n}
\eeq
where 
\beq
{\bf J}^2_k=\Tr \nu_k^2
\quad\quad
{\bf H}^{(0)}_k=2\sum_{l\neq k}\frac{1}{w_k-w_l} \Tr \nu_l\nu_k
\label{H0k}
\eeq
According to the general discussion above, the quantization of these operators
is provided by replacing $\nu_k$ at each puncture with the corresponding spin operators.

  Analogously, in the torus case, the Higgs field is
    \beq
    \tilde\Phi
     =     \left( \sum_k \nu_k \frac{\vartheta_1'(z-w_k)}{\vartheta_1(z-w_k)} + 2\pi i p \right) \frac{dz}{2\pi i}
    \eeq
  from which it follows that
    \beq
    \Tr \Phi^2
     =     \sum_k \left({\cal P}(z-w_k) {\bf J}^2_k + \frac{\vartheta_1'(z-w_k)}{\vartheta_1(z-w_k)} {\bf H}^{(1)}_k + {\bf H}^{(1)}_0\right)
           \left(\frac{dz}{2\pi i}\right)^2,
           \label{toro}
    \eeq
  where\footnote{Notice that with respect to (4.10) in \cite{gaw}, in ${\bf H}^{(1)}_0$ we find also a 
  term proportional to $\eta_1$, which will reveal to be crucial in comparing with the quantum Seiberg-Witten curve. 
  This term was invisible to the authors of \cite{gaw}, being the absolute 
  normalization of the differential which they admittedly do not check.} 
    \bea
    {\bf H}_k^{(1)}
    &=&    2\sum_{l\neq k}\Tr \nu_k\nu_l\frac{\vartheta_1'(w_k-w_l)}{\vartheta_1(w_k-w_l)}+4\pi i \Tr\nu_k p,
           \nonumber \\
    {\bf H}_0^{(1)}
    &=&  - 4\pi^2\Tr p^2 -\eta_1 \sum_k {\bf J}^2_k +\frac{1}{2}\sum_{k,l; k\neq l}
           \Tr\nu_k\nu_l\frac{\vartheta_1''(w_k-w_l)}{\vartheta_1(w_k-w_l)}.
    \eea
  See Appendix A for the definition of the theta functions.
  In (\ref{toro}), ${\cal P}$ is the Weierstrass ${\cal P}$-function. 
  To obtain (\ref{toro}) we used the identity (\ref{pformula}) 
  relating the Weierstrass ${\cal P}$-function and its primitive $\zeta'(z)=-{\cal P}(z)$. 
  
  In what follows, we will see the appearing of the above Hitchin Hamiltonians
  via the generalized beta-deformed matrix model.

\section{Matrix model: genus zero}
\label{sec:sphere}
  The AGT relation associated with a sphere is the one between the Nekrasov partition function
  of $\CN=2$ superconformal $SU(2)^{n-3}$ linear quiver gauge theory 
  and the $n$-point conformal block on the sphere.
  Both of them are specified by the marking of $\mathcal{C}_{0, n}$
  We can obtain the beta-deformed matrix model starting 
  from the Dotsenko-Fateev integral representation of the conformal block \cite{DF} as follows.
  (See \cite{DV, EM2}).
  In terms of the free field $\phi(z)$ (whose OPE is $\phi(z) \phi(\omega) \sim - \frac{1}{2} \log (z - \omega)$),
  the $n$-point conformal block is described by inserting the screening operators
    \bea
    Z^{{\mathcal C}_{0, n}}
     =     \left< \left( \int d \lambda_I :e^{2b \phi(\lambda_I)} : \right)^N ~
           \prod_{k=0}^{n-1} V_{m_k}(w_k) \right>_{{\rm free~on~}\mathcal{C}_0},
           \label{freefieldC0n}
    \eea
  where the vertex operator $V_{m_k}(w_k)$ is given by $:e^{2 m_k \phi(w_k)}:$.
  The momentum conservation condition relates the external momenta and the number of integrals as 
  $\sum_{k=0}^{n-1} m_k = b N - Q$.
  By evaluating the OPEs, it is easy to obtain
    \bea
    Z^{{\mathcal C}_{0, n}}
     =     C(m_k, w_k) \widetilde{Z}^{{\mathcal C}_{0, n}}
     \equiv
           e^{F^{{\mathcal C}_{0, n}}/g_s^2},
           \label{ZC0n}
    \eea
  where $\widetilde{Z}^{{\mathcal C}_{0, n}}$ is the beta-deformation of one matrix model
    \bea
    \widetilde{Z}^{{\mathcal C}_{0, n}}
     =     \int \prod_{I=1}^N d \lambda_I \prod_{I < J} (\lambda_I - \lambda_J)^{-2b^2} 
           e^{\frac{b}{g_s} \sum_I W(\lambda_I)}
     \equiv
           e^{\tilde{F}^{{\mathcal C}_{0, n}}/g_s^2},
           \label{betamatrix}
    \eea
  and 
    \bea
    W(z)
     =     \sum_{k=0}^{n-2} 2m_k \log (z - w_k), ~~~~
    C(m_k, w_k)
     =     \prod_{k < \ell \leq n-2} (w_k - w_\ell)^{- \frac{2 m_k m_\ell}{g_s^2}}.
           \label{potentialC0n}
    \eea
  We have introduced the parameter $g_s$ by rescaling $m_k \rightarrow \frac{m_k}{g_s}$.
  We relate $m_k$ with the mass parameters of the gauge theory.
  Also, we have chosen three insertion points as $w_0 = 0$, $w_1 = 1$ and $w_{n-1} = \infty$.
  The remaining parameters are identified with the gauge theory coupling constants 
  $q_i =e^{2 \pi i \tau_i}$ ($i=1, \ldots, n-3$) as follows:
    \bea
    w_2
     =     q_1, ~~
    w_3
     =     q_1 q_2, ~~ \ldots, ~~
    w_{n-2}
     =     q_1 q_2 \ldots q_{n-3}.
           \label{moduliidentificationC0n}
    \eea
  While the dependence on $m_{n-1}$ disappeared in the potential, 
  this is recovered by the momentum conservation condition
    \bea
    \sum_{k=0}^{n-2} m_k + m_{n-1}
     =     b g_s N - g_s Q.
           \label{momentumconservationC0n}
    \eea
  We will refer to $F_m$ as free energy.
  
  The identification of the parameter $b$ with the Nekrasov's deformation parameters is given by 
    \bea
    \epsilon_1
     =     b g_s, ~~~
    \epsilon_2
     =     \frac{g_s}{b}.
    \eea
  Note that, in the case of $b=i$ ($c=1$), this reduces to the usual hermitian matrix model
  and this case corresponds to the self-dual background $\epsilon_1 = - \epsilon_2$. 
  
  Here we define the resolvent of the matrix model as
    \bea
    R(z_1, \ldots, z_k)
     =     (b g_s)^k \sum_{I_1} \frac{1}{z_1 - \lambda_{I_1}} .... \sum_{I_k} \frac{1}{z_k - \lambda_{I_k}}.
           \label{resolventsphere}
    \eea
  For $k=1$, this reduces to the usual resolvent.

\subsection{Wave-function and conformal block}
  In the following, we mainly concentrate on the limit where $\epsilon_2 \rightarrow 0$ 
  with $\epsilon_1$ and the other parameters keeping fixed.
  In other words, the limit is $b \rightarrow \infty$ and $g_s \rightarrow 0$ with $b g_s$ and $N$ keeping finite.
  This is the limit by Nekrasov and Shatashvili \cite{NS}.
  In \cite{Teschner:2010je, MT}, it was shown that the the conformal blocks on a sphere
  with the additional insertion of the degenerate fields $V_{\frac{1}{2b}}(z) = e^{- \frac{\phi(z)}{b}}$ 
  capture the quantization of the integrable systems.
  
  In this limit, the beta-deformed partition function can be written as
  $\int \prod_{I=1}^N d \lambda_I \exp (- \frac{1}{\epsilon_2} \widetilde{W})$
  where
    \bea
    \widetilde{W} 
     =     \sum_I W(\lambda_I) + 2 \epsilon_1 \sum_{I<J} \log (\lambda_I - \lambda_J).
    \eea
  Thus the leading order part of the free energy can be obtained from the value of the critical points 
  which solve the equations of motion:
    \bea
    W'(\lambda_I) + 2 \epsilon_1 \sum_{J \neq I} \frac{1}{\lambda_I - \lambda_J}
     =     0.
    \eea
  We note that these two terms are of the same order in the limit 
  because $N$ and $\epsilon_1$ are kept finite.
  
  In this section, we will show that 
  under the identification the beta-deformed matrix model $Z^{{\mathcal C}_{0, n}}$ with the $n$-point conformal block, 
  the integral representation of the degenerate conformal block can be written in terms of the resolvent of 
  the original matrix model (\ref{resolventsphere}), in the $\epsilon_2 \rightarrow 0$ limit.
  More explicitly, we will show
    \bea
    \frac{Z^{{\mathcal C}_{0,n+n}}_{deg}}{Z^{{\mathcal C}_{0,n}}} (z_1, z_2, \ldots, z_n)
     =     \prod_{i=1}^n \Psi_i(z_i), ~~~
    \Psi_i (z_i)
     =     \exp \left( \frac{1}{\epsilon_1} \int^{z_i} x(z') dz' \right),
           \label{PsiZ5}
    \eea
  in the $\epsilon_2 \rightarrow 0$ limit, 
  where $Z^{{\mathcal C}_{0,n+n}}_{deg}$ is the matrix model partition function 
  corresponding to the $2n$-point conformal block with $n$ degenerate fields inserted at $z = z_i$.
  This property of ``separation of variables" agrees with the corresponding result of the Virasoro conformal block 
  as in \cite{sara, Teschner:2010je}.
  The differential $x(z) dz$ is identified with the ``quantized" Seiberg-Witten differential which
  is given, in terms of matrix model language, by 
    \bea
    x(z)
     =     \left< R(z) \right> - \frac{W'(z)}{2},
           \label{x}
    \eea
  where $\left< \ldots \right> = \frac{1}{Z^{{\mathcal C}_{g,n}}} \int \prod d \lambda 
  \prod (\lambda_I - \lambda_J)^{-2b^2} e^{b \sum W/g_s} \ldots$.
  This relation (\ref{PsiZ5}) is a simple generalization of 
  the one obtained in \cite{MMM} for the single degenerate field insertion.
  
  First of all, we consider a more generic expression: the ($n+\ell$)-point conformal block 
  where $\ell$ degenerate fields are inserted
    \bea
    Z^{{\mathcal C}_{0,n+\ell}}_{deg}
    &=&    \left< \prod_{i=1}^\ell V_{\frac{1}{2b}}(z_i) \left( \int d \lambda e^{2b \phi(\lambda)} \right)^N ~
           \prod_{k=0}^{n-1} V_{\frac{m_k}{g_s}}(w_k) \right>
           \nonumber \\
    &=&    \prod_{i<j} (z_i - z_j)^{- \frac{1}{2 b^2}} 
           \prod_{0 \leq k < \ell \leq n-2} (w_k - w_\ell)^{- \frac{2 m_k m_\ell}{g_s^2}} 
           \prod_{i=1}^\ell \prod_{k=0}^{n-2} (z_i - w_k)^{- \frac{m_k}{b g_s}}
           \nonumber \\
    & &    ~~~~~~\times
           \int \prod_{I=1}^N d \lambda_I \prod_{I<J} (\lambda_I - \lambda_J)^{- 2b^2}
           \prod_{I} \prod_{k=0}^{n-2} (\lambda_I - w_k)^{\frac{2 b m_k}{g_s}} \prod_{i=1}^\ell (z_i - \lambda_I),
           \label{degenerateC0n}
    \eea
  where the potential $W(z)$ is the same as (\ref{potentialC0n}).
  We have taken $w_{n-1}$ to infinity and omitted the factor including this, as we have done above.
  The momentum conservation is however modified by the degenerate field insertion as
    \bea
    \sum_{k=0}^{n-1} m_k + \frac{\ell g_s}{2 b} 
     =     b g_s N - g_s Q.
           \label{momentumC0n+ell}
    \eea
  By dividing by $Z^{{\mathcal C}_{0,n}}$ and taking a log, we obtain
    \bea
    \log \frac{Z^{{\mathcal C}_{0,n+\ell}}_{deg}}{Z^{{\mathcal C}_{0,n}}}
     =   - \frac{1}{2 b^2} \sum_{i<j} \log (z_i - z_j) 
         - \sum_i \frac{W(z_i)}{2 b g_s} + \log \left< \prod_{i, I} (z_i - \lambda_I) \right>.
           \label{logratio}
    \eea
  Notice that the expectation value is defined as above, 
  but with the modified momentum conservation (\ref{momentumC0n+ell}).
  By defining $e^L = \prod_{i, I} (z_i - \lambda_I)$, 
  we notice that
    \bea
    L
     =     \sum_{i, I} \log (z_i - \lambda_I)
     =     \sum_{i, I} \int^{z_i} \frac{d z_i'}{z_i' - \lambda_I},
           \label{integral}
    \eea
  where we have ignored irrelevant terms due to the end points of the integrations. 
  Then, we use that the expectation value of $e^L$ can be written as 
  $\log \left< e^L \right> = \sum_{k=1}^\infty \frac{1}{k!} \left< L^k \right>_{conn}$ \cite{MMM}, 
  where $\left< \ldots \right>_{conn}$ means the connected part of the correlator,  
  $\langle L^2 \rangle_{conn} = \langle L^2 \rangle - \langle L \rangle^2$, etc.
  Thus, (\ref{logratio}) can be expressed, by using (\ref{integral}), as
    \bea
    \log \frac{Z^{{\mathcal C}_{0,n+\ell}}_{deg}}{Z^{{\mathcal C}_{0,n}}}
     =   - \sum_i \frac{W(z_i)}{2 \epsilon_1} + \sum_{k=1}^\infty \frac{1}{k!} 
           \left< \left( \sum_{i,I} \int^{z_i} \frac{dz'}{z' - \lambda_I} \right)^k \right>_{conn}.
    \eea
  
  We will now consider the limit where $\epsilon_2 \rightarrow 0$. 
  In this limit, the terms with $k>1$ are subleading contributions compared with the $k=1$ terms 
  since the connected part of the expectation value can be ignored in this limit.
  Thus, we obtain
    \bea
    \log \frac{Z^{{\mathcal C}_{0,n+\ell}}_{deg}}{Z^{{\mathcal C}_{0,n}}}
     \rightarrow
            \frac{1}{\epsilon_1} \sum_i \int^{z_i} x(z') dz',
    \eea
  where we have used (\ref{x}).
  Thus, by setting $\ell = n$, we have obtained (\ref{PsiZ5}).
  This indicates that the properties of the conformal block with degenerate field insertions
  are build in the resolvent of the matrix model in the $\epsilon_2 \rightarrow 0$ limit.

\subsection{Loop equations}
  The argument in the previous section shows that 
  the relation with the integrable system can be seen by analyzing the resolvent,
  in particular, the loop equations.
  Thus, we derive it here with finite $\beta$.
  First of all, we keep the potential arbitrary and obtain
    \bea
    0
    &=&    \frac{1}{\widetilde{Z}^{{\mathcal C}_{0, n}}} 
           \int \prod_{I=1}^N d \lambda_I \sum_K \frac{\partial}{\partial \lambda_K}
           \left[ \frac{1}{z - \lambda_K} \prod_{I < J} (\lambda_I - \lambda_J)^{- 2b^2} 
           e^{\frac{b}{g_s} \sum_I W(\lambda_I)} \right]
           \nonumber \\
    &=&  - \frac{1}{g_s^2} \langle R(z,z) \rangle - \frac{b + \frac{1}{b}}{g_s} \langle R(z)' \rangle
         + \frac{1}{g_s^2} W'(z) \langle R(z) \rangle - \frac{f(z)}{g_s^2},
           \label{SD}
    \eea
  where $R'$ is the $z$-derivative of the resolvent and 
  we have defined 
    \bea
    f(z)
     =     b g_s \left< \sum_I \frac{W'(z)- W'(\lambda_I)}{z - \lambda_I} \right>.
    \eea
  The expectation value is defined as the matrix model average.
  By multiplying  (\ref{SD}) by $g_s^2$, we obtain
    \bea
    0
     =   - \langle R(z,z) \rangle - (\epsilon_1 + \epsilon_2) \langle R(z)' \rangle
         + \langle R(z) \rangle W'(z) - f(z).
           \label{loopequation}
    \eea
  In the case of the hermitian matrix model $b=i$, the second term vanishes and 
  the equation reduces to the well-known one.

  Let us then analyze $f(z)$ by specifying the potential to the Penner type one (\ref{potentialC0n}).
  In this case, 
    \bea
    f(z)
     =     \sum_{k=0}^{n-2} \frac{c_k}{z - w_k},
    \eea
  where for $k \geq 2$
    \bea
    c_k
     =   - b g_s \left< \sum_I \frac{2 m_k}{\lambda_I - w_k} \right>
     =     g_s^2 \frac{\partial \log \widetilde{Z}^{{\mathcal C}_{0, n}}}{\partial w_k}
     =     \frac{\partial \widetilde{F}_m}{\partial w_k}.
           \label{ck}
    \eea
  While we cannot write $c_0$ and $c_1$ as above because we have chosen $w_0 =0$ and $w_1 = 1$, 
  we can see that they are written in terms of $c_k$ with $k \geq 2$.
  First of all, due to the equations of motion: $\left< \sum_I W'(\lambda_I) \right> = 0$, 
  the sum of $c_k$ is constrained to vanish $\sum_{k=0}^{n-2} c_k = 0$.
  In order to find another constraint, we consider the asymptotic at large $z$ of the loop equation.
  The asymptotic of the resolvent is
  $\langle R(z) \rangle  \sim \frac{b g_s N}{z}$, so that
  the leading terms at large $z$ in the loop equations satisfy
    \bea
    - (b g_s N)^2 + (\epsilon_1 + \epsilon_2) b g_s N + b g_s N \sum_{k=0}^{n-2} 2 m_k - \sum_{k=0}^{n-2} w_k c_k
     =     0.
    \eea
  The leading term of order $1/z$ in $f(z)$ vanishes via the first constraint.
  Thus, we obtain
    \bea
    \sum_{k=0}^{n-2} w_k c_k
     =     \left( \sum_{k=0}^{n-2} m_k + m_{n-1} + \frac{n g_s}{2 b} + g_s Q \right)
           \left( \sum_{k=0}^{n-2} m_k - m_{n-1} + \frac{n g_s}{2 b} \right)
     \equiv
           M^2,
           \label{crelation2}
    \eea
  where we have used the momentum conservation\footnote{We are using the modified momentum conservation to 
                                                        apply this to the argument in the previous section.
                                                        However, the difference will disappear 
                                                        in the $\epsilon_2 \rightarrow 0$ limit.} 
  (\ref{momentumC0n+ell}) with $\ell = n$.
  Therefore, $c_0$ and $c_1$ can be written in terms of $c_k$ (\ref{ck}).
  These constraints are related to the Virasoro constraints \cite{BPZ}.

\subsubsection*{$\epsilon_2 \rightarrow 0$ limit}
  As above, in the $\epsilon_2 \rightarrow 0$ limit, 
  the connected part of (\ref{resolventsphere}) can be ignored in this limit: 
  $\langle R(z,z) \rangle \rightarrow \langle R(z) \rangle^2$.
  Taking this into account, the loop equation (\ref{loopequation}) becomes
    \bea
    0
     =   - \langle \tilde{R}(z) \rangle^2 - \epsilon_1 \langle \tilde{R}(z)' \rangle
         + \langle \tilde{R}(z) \rangle W'(z) - \tilde{f}(z),
    \eea
  where $\tilde{R}$ and $\tilde{f}$ are $R|_{\epsilon_2 \rightarrow 0}$ and $f|_{\epsilon_2 \rightarrow 0}$ respectively.
  In the following, we will omit the tildes of $R$ and $f$.
  Then, in terms of $x = \langle R(z) \rangle- \frac{W'(z)}{2}$, the equation becomes
    \bea
    0
     =   - x^2 - \epsilon_1 x' + U(z),
    \eea
  where
    \bea
    U(z)
     =     \frac{W'(z)^2}{4} - \frac{\epsilon_1}{2} W^{''}(z) - f(z).
           \label{U}
    \eea
  This equation is similar to the one obtained in \cite{EO}.
  It is easy to see that this can be written as the Schr\"odinger-type equation:
    \bea
    0
     =   - \epsilon_1^2 \frac{\partial^2}{\partial z^2} \Psi(z) + U(z) \Psi(z),
           \label{diffC0n}
    \eea
  where $\Psi (z)$ is defined in (\ref{PsiZ5}).
  
  The above argument is applicable for an arbitrary potential $W(z)$.
  Here we return to the Penner-type one (\ref{potentialC0n}) and see the relation with 
  the Gaudin Hamiltonian\footnote{A relation with Gaudin system at finite $\epsilon_2$ has been noticed also in \cite{EO}}.
  (\ref{U}) becomes in this case
    \bea
    U(z)
     =     \sum_{k=0}^{n-2} \frac{m_k (m_k + \epsilon_1)}{(z - w_k)^2}
         + \sum_k \frac{H_k}{z - w_k}
         - \sum_{k=0}^{n-2} \frac{c_k}{z-w_k}
           \label{UC0n}
    \eea
  where
    \bea
    H_k
     =     \sum_{\ell(\neq k)} \frac{2 m_k m_\ell}{w_k - w_\ell}.
    \eea
  $U(z)$ is the vacuum expectation value of (\ref{HamiltonianC0n}).
  In particular, notice that the residue of the quadratic pole in (\ref{UC0n}) corresponds to the 
  eigenvalue of ${\bf J}_k^2$ quantized in $\epsilon_1$ units
  and that $H_k - c_k$ are the vacuum energies of the quantum Hamiltonians ${\bf H}^{(0)}_k$ (\ref{H0k}).

  Let us rewrite $c_k$ in terms of the gauge theory variables.
  Since the moduli of the sphere are related with the gauge coupling constants of the linear quiver gauge theory
  as in (\ref{moduliidentificationC0n}), 
  the derivatives with respect to $w_k$ can be written as 
  $2 \pi i w_k \frac{\partial}{\partial w_k} = \frac{\partial}{\partial \tau_{k-1}} - \frac{\partial}{\partial \tau_k}$,
  where the second term vanished when $k=n-2$.
  Therefore, for $k=2, \ldots, n-2$, by using (\ref{ck}), we obtain
    \bea
    c_k
     =     \frac{1}{2 \pi i w_k} 
           \left( \frac{\partial \tilde{F}^{\mathcal{C}_{0,n}}}{\partial \tau_{k-1}}
         - \frac{\partial \tilde{F}^{\mathcal{C}_{0,n}}}{\partial \tau_k} \right)
     =     \frac{1}{2 \pi i w_k} 
           \left( u_{k-1} - u_{k} \right).
    \eea
  where $u_k$ are closely related with the gauge theory variables $\langle \tr \phi^2_k \rangle$,
  $\phi_k$ being the vector multiplet scalar of the $k$-th gauge group.
  Indeed, supposing that the free energy $F^{\mathcal{C}_{0,n}}$ is identified 
  with the prepotential (with $\epsilon_1$ and $\epsilon_2$) of the gauge theory 
  (this has indeed been checked for $n=4$ in \cite{MMS,MMS2,IO, MMMorozov} in some orders in the moduli),
  we can use the $\epsilon$-deformed version \cite{Flume:2004rp, Fucito:2005wc, cicca} 
  of the Matone relation \cite{Matone, STY, EY} to relate $u_k$ with $\langle \tr \phi^2_k \rangle$.
  Note that there is still difference between $\tilde{F}^{\mathcal{C}_{0,n}}$ and $F^{\mathcal{C}_{0,n}}$,
  we will explicitly consider this in an example below.
  Instead, for $c_0$ and $c_1$, we can use the two constraints derived above and obtain
    \bea
    c_0
    &=&    \sum_{k=2}^{n-2} (w_k - 1) c_k - M^2
     =     \frac{1}{2 \pi i} \sum_{k=2}^{n-2} \frac{w_k - 1}{w_k} (u_{k-1} - u_{k} ) - M^2 , 
           \nonumber \\
    c_1
    &=&  - \sum_{k=2}^{n-2} w_k c_k + M^2
     =   -  \frac{1}{2 \pi i} \sum_{k=2}^{n-2} (u_{k-1} - u_{k} ) + M^2.
    \eea
  
  We have obtained the differential equations for $\Psi = \prod_i \Psi_i (z_i)$
  each of which is (\ref{diffC0n}) satisfied by $\Psi_i (z_i)$.
  These are similar to the differential equations satisfied by the ($n+n$)-point Virasoro conformal block 
  where $n$ vertex operators are chosen to be degenerate $V(z) = e^{-\frac{\phi(z)}{b}}$, 
  as in \cite{ Teschner:2010je}.
  As an example, we will explicitly see in the subsequent section the differential equation for $n=4$ is
  the same as that obtained from the Virasoro conformal block.

\subsection{Sphere with four punctures}
  We now consider the case corresponding to a sphere with four puncture 
  where the matrix model potential is given by
    \bea
    W(z)
     =     \sum_{k=0}^2 2 m_k \log (z - w_k), ~~~~~
    w_0
     =     0, ~~
    w_1
     =     1, ~~
    w_2 
     =     q.
    \eea
  This corresponds to $SU(2)$ gauge theory with four flavors
  and the Gaudin model on a sphere with four punctures where the number of commuting Hamiltonian is just one.
  Thus we simply consider the wave-function $\Psi(z) = e^{\frac{1}{\epsilon_1} \int^z x dz'}$
  and the differential equation satisfied by it.
  
  In this case, $U(z)$ can be evaluated in terms of $c_2$ as
    \bea
    U(z)
    &=&    \sum_{k=0}^2 \frac{m_k^2 + \epsilon_1 m_k}{(z - w_k)^2} 
         + \frac{m_3^2 - \sum_{i=0}^2 m_i^2 + \epsilon_1 (m_3 - \sum_{k=0}^2 m_k)}{z (z - 1)}
           \nonumber \\
    & &    
         + \frac{- q (q - 1) c_2 + 2 (q m_1 m_2 + (q - 1)m_2 m_0)}{z (z - 1) (z - q)},
    \eea
  Let us relate this with the one obtained from the Virasoro conformal block.
  We consider the last line of $U(z)$.
  Let us recall the definition of the free energy (\ref{ZC0n}) and (\ref{betamatrix}).
  Since the difference of them is expressed 
  by $C = q^{-\frac{2 m_2 m_0}{g_s^2}} (1 - q)^{- \frac{m_1 m_2}{g_s^2}}$ in this case,
  the free energies are related by
    \bea
    F^{\mathcal{C}_{0,4}}
     =     \widetilde{F}^{\mathcal{C}_{0,4}} - 2 m_1 m_2 \log (1 - q) - 2 m_2 m_0 \log q.
    \eea
  Therefore, its derivative is
    \bea
    q (1 - q) \frac{\partial F^{\mathcal{C}_{0,4}}}{\partial q}
     =     q (1 - q) \frac{\partial \widetilde{F}^{\mathcal{C}_{0,4}}}{\partial q} + 2 q m_1 m_2 - 2 (1-q) m_2 m_0.
    \eea
  We notice that the right hand side is the numerator of the last line of $U(z)$.
  
  Here let us redefine the mass parameters as
    \bea
    \tilde{m}_0
     =     m_0 + \frac{\epsilon_1}{2}, ~~~
    \tilde{m}_3
     =     m_3 - \frac{\epsilon_1}{2}.
    \eea
  In this notation, $U$ can be written as
    \bea
    U(z)
    &=&    \frac{\tilde{m}_0^2 - \frac{\epsilon_1^2}{4}}{z^2} + \frac{m_1(m_1 + \epsilon_1)}{(z-1)^2}
         + \frac{m_2(m_2 + \epsilon_1)}{(z - q)^2}
           \nonumber \\
    & &  - \frac{- \tilde{m}_3^2 + \tilde{m}_0^2 + m_1(m_1 + \epsilon_1) + m_2(m_2 + \epsilon_1)}{z (z - 1)}
         + \frac{q (1 - q)}{z (z-1)(z - q)} \frac{\partial F}{\partial q}.
    \eea
  Note here that the first four terms are exactly the potential which considered in \cite{MT} $V(z)$.
  Also, the last term might correspond to the ``eigenvalue" in \cite{MT}.
  The Schr\"odinger-like equation becomes
    \bea
    - \epsilon_1^2 \frac{\partial^2}{\partial z^2} \Psi (z) + V(z) \Psi(z)
     =   - \frac{q (1 - q)}{z (z-1)(z - q)} \frac{\partial F}{\partial q} \Psi(z).
    \eea
  Note that this differential equation has also been derived in \cite{MMM}
  from the free field expression, the first line of (\ref{degenerateC0n}).

\section{Matrix model: genus one}
\label{sec:torus}
  In this section, we consider the matrix model corresponding to the conformal block on a torus with punctures
  \cite{DV, MY, MMS, BMTY}.
  We will derive the loop equations of the matrix model 
  and relate it with the differential equations of the corresponding Hitchin system.
  
  We consider the $n$-point conformal block on a torus whose integral description is
    \bea
    Z^{{\mathcal C}_{1,n}}
     =     e^{F^{{\mathcal C}_{1, n}}/g_s^2}
    &=&    \int \prod_I d \lambda_I \prod_{I<J} 
           \left[ \frac{\vartheta_1(\lambda_I - \lambda_J)}{\eta(\tau)^3} \right]^{-2b^2}
           \prod_I \prod_{k=1}^n \left[ \frac{\vartheta_1(\lambda_I - w_k)}{\eta(\tau)^3} \right]^{\frac{2 b m_k}{g_s}} 
           e^{\frac{4 \pi b p}{g_s} \sum_I \lambda_I} 
           \nonumber \\
    & &    ~~~\times
           \prod_{k < \ell} \left[ \frac{\vartheta_1(w_k - w_\ell)}{\eta(\tau)^3} \right]^{- \frac{2 m_k m_\ell}{g_s^2}} 
           e^{ - \frac{4 \pi p \sum_k m_k w_k}{g_s^2}}
           \nonumber \\
    &=&    C(w_k, m_k, p) \widetilde{Z}^{{\mathcal C}_{1,n}},
           \label{partitiontorusn}
    \eea
  where $C(w_k, m_k, p)$ is the $\lambda$ independent coefficient
    \bea
    C(w_k, m_k, p)
     =     \eta^{-3(-b^2 N (N-1) + \frac{2bN}{g_s} \sum_k m_k + \frac{2}{g_s^2} \sum_{k<\ell} m_k m_\ell)}
           \prod_{k < \ell} \vartheta_1(w_k - w_\ell)^{- \frac{2 m_k m_\ell}{g_s^2}} 
           e^{ - \frac{4 \pi p \sum_k m_k w_k}{g_s^2}}
           \label{C}
    \eea
  and thus,
    \bea
    \widetilde{Z}^{{\mathcal C}_{1,n}}
     =     e^{\widetilde{F}^{{\mathcal C}_{1, n}}/g_s^2}
     =     \int \prod_I d \lambda_I \prod_{I<J} 
           \vartheta_1(\lambda_I - \lambda_J)^{-2b^2} e^{\frac{b}{g_s} \sum_I W(\lambda_I)}
    \eea
  with the potential
    \bea
    W(z)
     =     \sum_{k=1}^n 2 m_k \log \vartheta_1(z - w_k) + 4 \pi p z.
           \label{potentialtorusn}
    \eea
  While we do not know the representation of this in terms of a matrix, we refer to this as (generalized) matrix model.
  As the matrix model in the case of the sphere, 
  this expression has been obtained \cite{MY, BMTY} from the free field expression, 
  similar to (\ref{freefieldC0n}) but on the torus, following from the Liouville correlator by the method in \cite{GL}.
  The parameters must satisfy the momentum conservation
    \bea
    \sum_{k=1}^n m_k
     =     b g_s N.
           \label{momentumtorusn}
    \eea
  See Appendix A for the definition of the elliptic theta functions.
  
  This matrix model is related with the $\CN=2$ elliptic $SU(2)$ quiver gauge theory \cite{Witten}
  which is obtained from two M5-branes on $\mathcal{C}_{1,n}$ with specifying its marking.
  The gauge theory coupling constants $q_i = e^{2 \pi i \tau_i}$ ($i=1, \ldots, n$) are identified 
  with the moduli of the torus as
    \bea
    e^{2 \pi i w_{k}}
     =     \prod_{i=k}^{n-1} q_{i}, ~~~
    q 
     \equiv 
           e^{2 \pi i \tau}
     =     \prod_{i=1}^{n} q_i.
    \eea
  The parameters $m_k$ are directly identified with the mass parameters of the bifundamentals.
  The remaining parameters of the matrix model, the filling fractions and momentum $p$ in the potential
  determine the Coulomb moduli of the gauge theory.
  
\subsection{Wave-function and conformal block}
\label{subsec:}
  As in the previous section, we can show the relation between the resolvent of this matrix model and 
  the conformal block with the degenerate fields $\Phi_{1,2} = e^{- \frac{1}{b} \phi(z)}$.
  The integral representation of the latter is
    \bea
    Z^{{\mathcal C}_{1,n+\ell}}_{deg}(z_i)
    &=&    C(w_k, m_k, p) \int \prod_I d \lambda_I \prod_{I<J} \vartheta_1(\lambda_I - \lambda_J)^{-2b^2}
           \prod_I \vartheta_1(\lambda_I - w_1)^{\frac{2 b m_1}{g_s}} 
           e^{\frac{4 \pi b p}{g_s} \sum_I \lambda_I}
           \\
    & &    \times \prod_{i=1}^\ell \left(
           \prod_{k} \left[ \frac{\vartheta_1(z_i - w_k)}{\eta(\tau)^3} \right]^{- \frac{m_k}{b g_s}} 
           \prod_I \left[ \frac{\vartheta_1(z_i - \lambda_I)}{\eta(\tau)^3} \right]
           e^{ - \frac{2 \pi p}{b g_s} z_i} \right)
           \prod_{i<j} \left[ \frac{\vartheta_1(z_i - z_j)}{\eta(\tau)^3} \right]^{- \frac{1}{2 b^2}} ,
           \nonumber 
    \eea
  where $z_i$ are the insertion points of the degenerate fields
  and the momentum conservation is slightly deformed from the original one to
    \bea
    \sum_{k=1}^n m_k + \frac{\ell g_s}{2b}
     =     b g_s N.
           \label{momentumtorusn+1}
    \eea
  As before, we consider $\log \frac{Z^{{\mathcal C}_{1,n+\ell}}_{deg}}{Z^{{\mathcal C}_{1,n}}}$.
  The main object we want to know is $\log \left< \prod_{I, i} \vartheta_1 (z_i - \lambda_I) \right>$,
  where the expectation value is defined in the same way as in the sphere case.
  Note that we are using the momentum conservation (\ref{momentumtorusn+1}).
  Then, if we define $e^L = \prod_{I, i} \vartheta_1 (z_i - \lambda_I)$, we can rewrite it as
  $L = \sum_{I, i} \int^{z_i} \frac{\vartheta_1'(z_i' - \lambda_I)}{\vartheta_1(z_i' - \lambda_I)} dz_i'$.
  By using the same argument as the sphere case: 
  $\log \left< e^L \right> = \sum_k \frac{1}{k!} \left< L^k \right>_{conn}$, 
  we therefore obtain
    \bea
    \log \frac{Z^{{\mathcal C}_{1,n+\ell}}_{deg}}{Z^{{\mathcal C}_{1,n}}}
    &=&  - \sum_i \frac{W(z_i)}{2 b g_s} - \frac{3\ell (\ell + 1)}{4b^2} \log \eta
         - \frac{1}{2b^2} \sum_{i<j} \log \vartheta(z_i - z_j)
           \nonumber \\
    & &    ~~~~~~~
         + \sum_{k = 1}^\infty \frac{1}{k!} \left< \left( \sum_{I, i} \int^{z_i} 
           \frac{\vartheta_1'(z_i' - \lambda_I)}{\vartheta_1(z_i' - \lambda_I)} dz_i' \right)^k \right>_{conn} .
           \label{logratioC1n+ell}
    \eea
  where we have used the momentum conservation in the second term.
  
  We introduce the deformation parameters and take the limit 
  where $\epsilon_2 \rightarrow 0$ while $\epsilon_1$ keeping fixed.
  In this limit, the path integral is dominated by the solutions of the equations of motion
    \bea
    W'(\lambda_I)
    - 2 b g_s \prod_{J \neq I} \frac{\vartheta_1'(\lambda_I - \lambda_J)}{\vartheta_1(\lambda_I - \lambda_J)}
     =     0,
           \label{eomtorusn}
    \eea
  as we have seen in the previous section, and 
  the connected part with $k \geq 2$ in (\ref{logratioC1n+ell}) is negligible.
  Thus we obtain
    \bea
    \log \frac{Z^{{\mathcal C}_{1,n+\ell}}_{deg}}{Z^{{\mathcal C}_{1,n}}}
    &\rightarrow&
           \frac{1}{\epsilon_1} \sum_{i} \int^{z_i} \left( \left< R(z_i') \right> - \frac{W'(z_i')}{2} \right) dz_i'
     \equiv
           \log \prod_i \Psi(z_i).
           \label{Psitorus}
    \eea
  where $R$ is an analog of the resolvent of the sphere case
    \bea
    R(z_1, \ldots, z_k)
     =     (b g_s)^k \sum_{I_1} \frac{\vartheta_1'(z_1 - \lambda_{I_1})}{\vartheta_1(z_1 - \lambda_{I_1})} \ldots 
           \sum_{I_k} \frac{\vartheta_1'(z_k - \lambda_{I_k})}{\vartheta_1(z_k - \lambda_{I_k})}.
           \label{resolventtorusn}
    \eea
  Thus, we have related $Z^{{\mathcal C}_{1,n+\ell}}_{deg}$ with the resolvent or $\Psi$ 
  at least in the $\epsilon_2 \rightarrow 0$ limit.
  
  One simple consequence which follows from the above formulas is about
  the monodromies of $\Psi$ or $Z^{{\mathcal C}_{1, n+1}}_{deg}$. 
  (For simplicity, we consider the $\ell=1$ case. 
  The generalization to $\ell \geq 2$ might be straightforward.)
  Along the $A$ cycle, $Z^{{\mathcal C}_{1,n+1}}_{deg}$ behaves as
    \bea
    Z^{{\mathcal C}_{1,n+1}}_{deg}(z + \pi)
     =     (-1)^{N - \frac{\sum_k m_k}{b g_s}} e^{- \frac{2 \pi^2}{b g_s} p} Z^{{\mathcal C}_{1,n+1}}_{deg}(z)
     =     (-1)^{\frac{1}{2 b^2}} e^{- \frac{2 \pi^2}{b g_s} p} Z^{{\mathcal C}_{1,n+1}}_{deg}(z),
           \label{monodromytorusA}
    \eea
  where we have used the momentum conservation.
  Similarly, we can evaluate the $B$-cycle monodromy
    \bea
    \frac{Z^{{\mathcal C}_{1,n+1}}_{deg} (z + \pi \tau)}{Z^{{\mathcal C}_{1,n}}}
    &=&    e^{- \frac{W(z)}{2 b g_s}} \eta^{- \frac{3}{2 b^2}} 
           \left< \prod_I \vartheta_1(z - \lambda_I) e^{2 i \sum_I \lambda_I} \right>
           \exp \left( - \frac{i z}{b^2} - \frac{\pi i \tau}{2 b^2} - \frac{2 \pi^2 \tau p}{b g_s} \right).
    \eea
  On the other hand, let us consider the shift of $p \rightarrow p - \frac{g_s}{2 \pi i b}$ 
  in $Z^{{\mathcal C}_{1, n+1}}_{deg}$, 
  which gives the additional factor $\exp \left(- \frac{i z}{b^2} + 2 i \sum_I \lambda_I \right)$
  in the integrals.
  Therefore, we obtain
    \bea
    Z^{{\mathcal C}_{1,n+1}}_{deg} (z + \pi \tau; p)
     =     Z^{{\mathcal C}_{1, n+1}}_{deg}(z; p - \frac{g_s}{2 \pi i b})
           e^{- \frac{2 \pi^2 \tau p}{b g_s} - \frac{\pi i \tau}{2 b^2}}.
    \eea
  We note that these are indeed the same monodromies 
  as those of the conformal block with the degenerate field \cite{AGGTV, MT}.
  
  We can further proceed to derive the special geometry ($\epsilon_1$-deformed Seiberg-Witten) relation
  for the resolvent.
  In the $\epsilon_2 \rightarrow 0$ limit, (\ref{Psitorus}) shows that 
  $Z^{{\mathcal C}_{1,n+1}}_{deg}$ can be expanded as
    \bea
    Z^{{\mathcal C}_{1,n+1}}_{deg} (z)
     =     \exp \left( \frac{\CF(\epsilon_1)}{\epsilon_1 \epsilon_2} + \frac{1}{\epsilon_1} \int^z x(z') dz'
         + \CO(\epsilon_2) \right),
    \eea
  where the first term is 
  $\CF(\epsilon_1) = \lim_{\epsilon_2 \rightarrow 0} F^{{\mathcal C}_{1,n}}$.
  The above $A$ and $B$ cycle monodromies indicate that for the second term
    \bea
    \oint_A x(z) dz
     =   - 2 \pi^2 p,~~~~
    \oint_B x(z) dz
     =   - 2 \pi^2 \tau p - \frac{1}{2 \pi i} \frac{\partial \CF(\epsilon_1)}{\partial p}.
           \label{specialtorus}
    \eea
  This corresponds to the Seiberg-Witten relation in the presence of the $\epsilon_1$ dependence \cite{MM, MM2}.
  For other independent cycles which correspond to the legs of the pants decomposition of the torus, 
  it is natural to expect that the similar relation is satisfied.

\subsection{Loop equations}
\label{subsec:loopC1n}
  Let us derive the loop equations of the matrix model for the torus.
  As discussed above, in order to relate with the conformal block, 
  we will use the momentum conservation (\ref{momentumtorusn+1}).
  From the Schwinger-Dyson equation for an arbitrary transformation 
  $\delta \lambda_K = \frac{\vartheta_1'(z - \lambda_K)}{\vartheta_1(z - \lambda_K)}$, we derive
    \bea
    0
    &=&    g_s^2 \left< \sum_I \left( \frac{\vartheta_1'(z - \lambda_I)}{\vartheta_1(z - \lambda_I)} \right)^2 \right>
         - g_s^2 \left< \sum_I \frac{\vartheta_1^{''}(z - \lambda_I)}{\vartheta_1(z - \lambda_I)} \right>
         + b g_s W'(z) \left< \sum_I \frac{\vartheta_1'(z - \lambda_I)}{\vartheta_1(z - \lambda_I)} \right>
           \nonumber \\
    & &  + t(z)
         - 2 b^2 g_s^2 \left< \sum_{I<J} \frac{\vartheta_1'(\lambda_I - \lambda_J)}{\vartheta_1(\lambda_I - \lambda_J)}
           \left( \frac{\vartheta_1'(z - \lambda_I)}{\vartheta_1(z - \lambda_I)}
         - \frac{\vartheta_1'(z - \lambda_J)}{\vartheta_1(z - \lambda_J)} \right) \right>,
           \label{SD1}
    \eea
  where we have multiplied the both sides by $g_s^2$ and defined
    \bea
    t(z)
     =     b g_s \left< \sum_I \frac{\vartheta_1'(z - \lambda_I)}{\vartheta_1(z - \lambda_I)} (W'(\lambda_I) - W'(z)) \right>.
    \eea
  We then use the formula (\ref{thetaformula}) to calculate the last term and, after some algebra, we obtain
    \bea
    0
    &=&  - \left< R(z,z) \right> - \left(b + \frac{1}{b} \right) g_s \left< R'(z) \right>
         + W'(z) \left< R(z) \right>
         + b^2 g_s^2 N \left< \sum_I \frac{\vartheta_1^{''}(z - \lambda_I)}{\vartheta_1(z - \lambda_I)} \right>
         + t(z)
           \nonumber \\
    & &  + b^2 g_s^2 \left< \sum_{I<J} \frac{\vartheta_1^{''}(\lambda_I - \lambda_J)}{\vartheta_1(\lambda_I - \lambda_J)} \right>
         + 3 b^2 g_s^2 \eta_1 N(N-1).
           \label{SD2}
    \eea
  This equation is valid for an arbitrary potential.
  
  From now on, let us consider the potential corresponding to the toric conformal block (\ref{potentialtorusn}).
  In this case, $t(z)$ can be evaluated by using (\ref{thetaformula}) again as
    \bea
    t(z)
    &=&    2 b g_s \sum_{k=1}^n m_k \frac{\vartheta_1'(z - w_k)}{\vartheta_1(z - w_k)} 
           \left< \sum_I \frac{\vartheta_1'(\lambda_I - w_k)}{\vartheta_1(\lambda_I - w_k)} \right>
         - b g_s \sum_{k = 1}^n m_k \left< \frac{\vartheta_1^{''}(\lambda_I - w_k)}{\vartheta_1(\lambda_I - w_k)} \right>
           \nonumber \\
    & &  - b g_s N \sum_{k=1}^n m_k \frac{\vartheta_1^{''}(z - w_k)}{\vartheta_1(z - w_k)}
         - b g_s \sum_{k=1}^n m_k \left< \sum_I \frac{\vartheta_1^{''}(z - \lambda_I)}{\vartheta_1(z - \lambda_I)} \right>
         - 6 b g_s N \eta_1 \sum_{k=1}^n m_k.
    \eea
  By substituting this into (\ref{SD2}) and using the momentum conservation, we obtain
    \bea
    0
    &=&  - \left< R(z,z) \right> - \left(b + \frac{1}{b} \right) g_s \left< R'(z) \right>
         + W'(z) \left< R(z) \right> - 3 b g_s (N+1) \eta_1 \sum_k m_k
           \nonumber \\
    & &  + 2 b g_s \sum_{k=1}^n m_k \frac{\vartheta_1'(z - w_k)}{\vartheta_1(z - w_k)} 
           \left< \sum_I \frac{\vartheta_1'(\lambda_I - w_k)}{\vartheta_1(\lambda_I - w_k)} \right>
         - b g_s N \sum_k m_k \frac{\vartheta_1^{''}(z - w_k)}{\vartheta_1(z - w_k)}
           \nonumber \\
    & &  - b g_s \sum_k m_k \left< \sum_I \frac{\vartheta_1^{''}(\lambda_I - w_k)}{\vartheta_1(\lambda_I - w_k)} \right>
         + b^2 g_s^2 \left< \sum_{I<J} \frac{\vartheta_1^{''}(\lambda_I - \lambda_J)}{\vartheta_1(\lambda_I - \lambda_J)} 
           \right>
           \nonumber \\
    & &  + \frac{g_s^2 \ell}{2} \left< \sum_I \frac{\vartheta_1^{''}(z-\lambda_I)}{\vartheta_1(z-\lambda_I)} \right>
         + \frac{3 \ell}{2} g_s^2 \eta_1 (N-1).
           \label{SD3}
    \eea
  where the last two terms come from the deformation of the momentum conservation.
  Let us note that the $\tau$-derivative of the partition function $\widetilde{Z}^{{\mathcal C}_{1,n}}$
    \bea
    \frac{4 g_s^2}{\widetilde{Z}^{{\mathcal C}_{1,n}}} 
    \frac{\partial \widetilde{Z}^{{\mathcal C}_{1,n}}}{\partial \ln q}
     =   - b g_s \sum_k m_k \left< \sum_I \frac{\vartheta_1^{''}(\lambda_I - w_k)}{\vartheta_1(\lambda_I - w_k)} \right>
         + b^2 g_s^2 \left< \sum_{I<J} \frac{\vartheta_1^{''}(\lambda_I - \lambda_J)}{\vartheta_1(\lambda_I - \lambda_J)}
           \right> ,
    \eea
  where we have used $-8 \frac{\partial}{\partial \ln q} \vartheta_1(a) = \vartheta_1^{''}(a)$.
  This is the third line in (\ref{SD3}).
  Similarly, the derivatives with respect to $w_k$ produce the first term in the second line of (\ref{SD3}):
    \bea
    - \frac{g_s^2}{\widetilde{Z}^{{\mathcal C}_{1,n}}} 
    \frac{\partial \widetilde{Z}^{{\mathcal C}_{1,n}}}{\partial w_k}
     =       2 b g_s m_k \left< \sum_I \frac{\vartheta_1'(\lambda_I - w_k)}{\vartheta_1(\lambda_I - w_k)} \right>.
             \label{righthandside}
    \eea
  We will fix one of the moduli of the torus as $w_n=0$ below.
  In this case, the derivative with respect to $w_n$ is understood as the right hand side of (\ref{righthandside}).
  Putting all these together, we finally obtain
    \bea
    0
    &=&  - \left< R(z,z) \right> - \left(b + \frac{1}{b} \right) g_s \left< R'(z) \right>
         + W'(z) \left< R(z) \right> - 3 b g_s (N+1) \eta_1 \sum_k m_k
           \nonumber \\
    & &  - g_s^2 \sum_{k=1}^n \frac{\vartheta_1'(z - w_k)}{\vartheta_1(z - w_k)} 
           \frac{\partial \ln \widetilde{Z}^{{\mathcal C}_{1,n}}}{\partial w_k}
         - b g_s N \sum_k m_k \frac{\vartheta_1^{''}(z - w_k)}{\vartheta_1(z - w_k)}
         + 4 g_s^2 \frac{\partial \ln \widetilde{Z}^{{\mathcal C}_{1,n}}}{\partial \ln q}
           \nonumber \\
    & &  + \frac{g_s^2 \ell}{2} \left< \sum_I \frac{\theta_1^{''}(z-\lambda_I)}{\theta_1(z-\lambda_I)} \right>
         + \frac{3 \ell}{2} g_s^2 \eta_1 (N-1).
           \label{SD4}
    \eea

\subsubsection*{$\epsilon_2 \rightarrow 0$ limit}
  We consider the $\epsilon_2 \rightarrow 0$ limit.
  We can use under this limit the equations of motion (\ref{eomtorusn}).
  Furthermore, as discussed in the previous section, $\left< R(z, z) \right> \sim \left< R(z) \right>^2$.
  Also, the last two terms in (\ref{SD4}) disappears in this limit.
  Naively, the terms with the derivative of $\ln \widetilde{Z}^{{\mathcal C}_{1,n}}$ also
  disappear due to the factor $g_s^2 (= \epsilon_1 \epsilon_2)$.
  However, they does not because $\ln \widetilde{Z}^{{\mathcal C}_{1,n}}$ behaves 
  as $\widetilde{F}^{{\mathcal C}_{1,n}}/g_s^2$.
  Thus, the loop equation (\ref{SD4}) becomes
    \bea
    0
    &=&  - \left< R(z) \right>^2 \Big|_{\epsilon_2 \rightarrow 0}
         - \epsilon_1 \left< R'(z) \right>\Big|_{\epsilon_2 \rightarrow 0}
         + W'(z) \left< R(z) \right>\Big|_{\epsilon_2 \rightarrow 0} 
         - 3 \eta_1 (\sum_\ell m_\ell) (\sum_k m_k + \epsilon_1)
           \nonumber \\
    & &  - (\sum_\ell m_\ell) \sum_k m_k \frac{\vartheta_1^{''}(z - w_k)}{\vartheta_1(z - w_k)}
         - \sum_{k=1}^n \frac{\vartheta_1'(z - w_k)}{\vartheta_1(z - w_k)} 
           \frac{\partial \widetilde{F}^{{\mathcal C}_{1,n}}}{\partial w_k} \Big|_{\epsilon_2 \rightarrow 0}
         + 4 \frac{\partial \widetilde{F}^{{\mathcal C}_{1,n}}}{\partial \ln q} 
           \Big|_{\epsilon_2 \rightarrow 0}.
    \eea
  We will omit $\Big|_{\epsilon_2 \rightarrow 0}$ in what follows.
  After some algebra (by introducing $x = \left< R(z) \right> - W'/2$ and by using the formula of the theta function), 
  we obtain
    \bea
    0
    &=&  - x^2 - \epsilon_1 x' + \sum_k m_k (m_k + \epsilon_1) \CP(z - w_k)
           \nonumber \\
    & &  + \sum_k \frac{\vartheta_1'(z - w_k)}{\vartheta_1(z - w_k)} 
           \left( H_k - \frac{\partial \widetilde{F}^{{\mathcal C}_{1,n}}}{\partial w_k} \right)
         + H_0 +  4 \frac{\partial \widetilde{F}^{{\mathcal C}_{1,n}}}{\partial \ln q},
           \label{SD5}
    \eea
  where $\CP(z)$ is the Weierstrass elliptic function (\ref{pe}) and 
    \bea
    H_k
    &=&    4 \pi p m_k + 2 \sum_{\ell(\neq k)} m_k m_\ell \frac{\vartheta_1'(w_k - w_\ell)}{\vartheta_1(w_k - w_\ell)},
           \nonumber \\
    H_0
    &=&    4 \pi^2 p^2 - \eta_1 \sum_k m_k (m_k + \epsilon_1)
         + \frac{1}{2} \sum_{k \neq \ell} m_k m_\ell \frac{\vartheta_1^{''}(w_k - w_\ell)}{\vartheta_1(w_k - w_\ell)}.
    \eea
  Therefore, we obtain that $H_k - \frac{\partial \widetilde{F}^{{\mathcal C}_{1,n}}}{\partial w_k}$ and 
  $H_0 +  4 \frac{\partial \widetilde{F}^{{\mathcal C}_{1,n}}}{\partial \ln q}$ are the vacuum energies
  of the quantum Hitchin Hamiltonians ${\bf H}^{(1)}_k$ and ${\bf H}^{(1)}_0$ in (\ref{toro}).
  This shows that this matrix model captures the quantization of the Hitchin system associated with the torus.
  
  We can write this equation in the form of the differential equation
  satisfied by the wave-function $\Psi(z) = e^{\frac{1}{\epsilon_1} \int^z x(z') dz'}$:
    \bea
    & &    \left( - \epsilon_1^2 \frac{\partial^2}{\partial z^2} + \sum_k m_k (m_k + \epsilon_1) \CP(z - w_k)
         + \sum_k \frac{\vartheta_1'(z - w_k)}{\vartheta_1(z - w_k)} 
           H_k + H_0 \right) \Psi(z)
           \nonumber \\
    & &    ~~~~~~~~
     =     \left( \sum_k \frac{\vartheta_1'(z - w_k)}{\vartheta_1(z - w_k)} 
           \frac{\partial \widetilde{F}^{{\mathcal C}_{1,n}}}{\partial w_k}
         - 4 \frac{\partial \widetilde{F}^{{\mathcal C}_{1,n}}}{\partial \ln q} \right) \Psi(z)
           \label{difftorus}
    \eea
  By considering $\prod_{i=1}^n \Psi(z_i)$, this satisfies the above differential equations for each $z_i$.
  This is related with the KZB equation \cite{B, EK, FW} as discussed in \cite{AT}
  (see also \cite{EOoguri}).
  
  It is easy to translate the terms in the right hand side in (\ref{difftorus}) to the gauge theory variables.
  By the identification of the moduli, the derivatives with respect to $w_k$ are written as
  $\frac{\partial}{\partial w_k} = \frac{\partial}{\partial \tau_k} - \frac{\partial}{\partial \tau_{k-1}}$
  (for $k=1, \ldots, n-1$), where $\tau_0 = \tau_n$, 
  and also $\frac{\partial}{\partial \ln q} = \frac{1}{2 \pi i} \frac{\partial}{\partial \tau_n}$.
  Therefore, we obtain
    \bea
    \frac{\partial \widetilde{F}^{{\mathcal C}_{1,n}}}{\partial w_k}
     =     u_k - u_{k-1},
    \eea
  for $k= 1, \ldots, n-1$,
  where we have defined $u_k = \frac{\partial \widetilde{F}^{{\mathcal C}_{1,n}}}{\partial \tau_k}$
  and $u_0 \equiv u_n$.
  For $k=n$, we calculate
    \bea
    \frac{\partial \widetilde{F}^{{\mathcal C}_{1,n}}}{\partial w_n}
     =   - \sum_{k=1}^{n-1} \frac{\partial \widetilde{F}^{{\mathcal C}_{1,n}}}{\partial w_k}
         + 4 \pi p \sum_{k=1}^n m_k
     =   - u_{n-1} + u_n + 4 \pi p \sum_{k=1}^n m_k
           \label{torusun}
    \eea
  where we have used the equations of motion and the momentum conservation.
  (We also ignored the term depending on $\epsilon_2$.)
  Also, for the derivative with respect to $q$ we obtain
    \bea
    - 4 \frac{\partial \widetilde{F}^{{\mathcal C}_{1,n}}}{\partial \ln q}
     =     - \frac{2}{\pi i} u_n.
    \eea
  As in the sphere case, these $u_k$ are related with $\langle \tr \phi^2_k \rangle$.
  Note however that there could be a difference of them since $u_k$ here are the derivatives of $\widetilde{F}$,
  which will be seen below.
  
  For completeness, let us rewrite the above equation in terms of the original partition function $Z^{{\mathcal C}_{1,n}}$.
  The difference between $Z^{{\mathcal C}_{1,n}}$ and $\widetilde{Z}^{{\mathcal C}_{1,n}}$
  is given by $C$ (\ref{C}), which gives rise to
    \bea
    4 \frac{\partial \widetilde{F}^{{\mathcal C}_{1,n}}}{\partial \ln q}
    &=&    4 \frac{\partial F^{{\mathcal C}_{1,n}}}{\partial \ln q}
         + 3 \sum_k m_k (m_k + \epsilon_1) \eta_1
         - \sum_{k<\ell} m_k m_\ell \frac{\vartheta_1^{''}(w_k - w_\ell)}{\vartheta_1(w_k - w_\ell)},
           \nonumber \\
    - \frac{\partial \widetilde{F}^{{\mathcal C}_{1,n}}}{\partial w_k}
    &=&  - \frac{\partial F^{{\mathcal C}_{1,n}}}{\partial w_k}
         - 2 \sum_{\ell(\neq k)} m_k m_\ell \frac{\vartheta_1'(w_k - w_\ell)}{\vartheta_1(w_k - w_\ell)}-4\pi p m_k,
           ~~~(k=1, \ldots, n-1)
           \nonumber \\
    - \frac{\partial \widetilde{F}^{{\mathcal C}_{1,n}}}{\partial w_n}
    &=&    \sum_{k=1}^{n-1} \frac{\partial F^{{\mathcal C}_{1,n}}}{\partial w_k}
         + \sum_{\ell (\neq n)} m_n m_\ell \frac{\vartheta_1'(w_\ell)}{\vartheta_1(w_\ell)}
         - 4 \pi p m_n,
    \eea
  where we have used (\ref{torusun}).
  Thus, we obtain from (\ref{SD5})
    \bea
    0
    &=&  - x^2 - \epsilon_1 x' + \sum_k m_k (m_k + \epsilon_1) \CP(z - w_k)
         + 2 \sum_k m_k (m_k + \epsilon_1) \eta_1
           \label{SD6} \\
    & &  - \sum_{k=1}^{n-1} \frac{\vartheta_1'(z - w_k)}{\vartheta_1(z - w_k)} 
           \frac{\partial F^{{\mathcal C}_{1,n}}}{\partial w_k}
         + \frac{\vartheta_1'(z)}{\vartheta_1(z)} 
           \left( \sum_{k=1}^{n-1} \frac{\partial F^{{\mathcal C}_{1,n}}}{\partial w_k}
         + \sum_{\ell (\neq n)} m_n m_\ell \frac{\vartheta_1'(w_\ell)}{\vartheta_1(w_\ell)} \right)
         + 4 \pi^2 p^2 +  4 \frac{\partial F^{{\mathcal C}_{1,n}}}{\partial \ln q}.
           \nonumber
    \eea

\subsection{One-punctured torus}
\label{subsec:}
  In the $n=1$ case, we can see the relation with the elliptic Calogero-Moser model.
  We will take $w_1 = 0$.
  The potential is
    \bea
    W(z)
     =     2 m_1 \log \vartheta_1(z) + 4 \pi p z.
           \label{potentialtorus1}
    \eea
  In this case, it is easy to calculate the loop equation (\ref{SD6}):
    \bea
    0
    &=&  - x(z)^2 - \epsilon_1 x'(z) + m_1 (m_1 + \epsilon_1) \CP(z) - 4 u(\epsilon_1),
           \label{SD7}
    \eea
  where 
    \bea
    u(\epsilon_1)
    &=&  - \pi^2 p^2
         - g_s^2 \frac{\partial \ln Z^{{\mathcal C}_{1,1}}}{\partial \ln q}
         - \frac{m_1 (m_1 + \epsilon_1) \eta_1}{2}
           \nonumber \\
    &=&  - \pi^2 p^2
         + \frac{\partial}{\partial \ln q} \left( \CF^{{\mathcal C}_{1,1}}
         - 2 m_1 (m_1 + \epsilon_1) \ln \eta \right),
           \label{utorusn}
    \eea
  where we have used that $\eta_1 = 4 \frac{\partial \ln \eta}{\partial \ln q}$ 
  (See Appendix A) and defined the free energy as 
  $\CF^{{\mathcal C}_{1,1}} = \lim_{\epsilon_2 \rightarrow 0} F^{{\mathcal C}_{1,1}}$.
  Note that the free energy is the one evaluated in the $\epsilon_2 \rightarrow 0$ limit.
Eq.(\ref{utorusn}) is the $\epsilon_1$-deformed version of the relation between $\langle\tr\phi^2\rangle$
and the Coulomb modulus $u$ \cite{Fucito:2005wc} which coincides with the one found in \cite{Fucito:2011pn}.
  
  By introducing the ``wave-function" $\Psi = e^{\frac{1}{\epsilon_1} \int^z dz' x(z')}$, we finally obtain
    \bea
    \left[- \epsilon_1^2 \frac{\partial^2}{\partial z^2} + m_1 (m_1 + \epsilon_1) \CP(z) \right] \Psi 
     =     4u \Psi.
           \label{CM}
    \eea
  The left hand side is the Calogero-Moser Hamiltonian and 
  $u$ in the right hand side can be considered as the eigenvalue of the Hamiltonian.
  Note that similar equations have been derived from the Virasoro conformal block \cite{MT} and 
  affine $s\hat{l}_2$ conformal block \cite{AT}.
  Indeed, by assuming the equivalence of the partition function of the matrix model $Z^{{\mathcal C}_{1,1}}$ 
  with the one-point conformal block on a torus,
  (\ref{CM}) becomes the exactly same equation as the one obtained from the conformal block with the degenerate field.
  (See Section 3.1.2 in \cite{MT}. The identification of the parameter is $a = i \pi p$.)
  
  We also emphasize that the differential $x dz$ in the wave-function satisfies
  the special geometry relation (\ref{specialtorus}).
  This is equivalent to the proposal in \cite{MM} stating that the $\epsilon_1$-deformed prepotential
  can be obtained from the $\epsilon_1$-deformed special geometry relation
  for the $\CN=2^*$ theory, by using the same argument as in \cite{MT}. 
  
\subsubsection{Large $N$ limit and prepotential}
  Before going to next, let us consider the loop equation in the large $N$ limit
  which can be obtain by taking $\epsilon_1 \rightarrow 0$ further in (\ref{SD7}):
    \bea
    x^2
    &=&    m^2 \CP(z) - 4 u(\epsilon_1 = 0),
    \eea
  which is the Seiberg-Witten curve of the $\CN=2^*$ gauge theory \cite{DW}.
  Indeed, the parameter $u$ can be written as
    \bea
    u (\epsilon_1 = 0)
     =   - \pi^2 p^2 + \frac{\partial}{\partial \ln q} \left( \CF^{{\mathcal C}_{1,1}}_0
         - 2 m_1^2  \ln \eta \right).
    \eea
  where $\CF^{{\mathcal C}_{1,1}}_0$ is the leading contribution of the full free energy 
  in the limit where $\epsilon_{1,2} \rightarrow 0$.
  The first term corresponds the classical contribution to the prepotential.
  The last term denotes the shift of the Coulomb moduli parameter from the value of the physical expectation value
  $\langle \Tr \phi^2 \rangle$ \cite{SW, Dorey, Fucito:2005wc}.

  Indeed, we can be more precise.
  Under the identification $i \pi p$ with the vev of the vector multiplet scalar $a$,
  it is easy to show from (\ref{monodromytorusA}) that
    \bea
    a
     =     i \pi p
     =     \frac{1}{2 \pi i} \int_0^\pi x dz.
    \eea
  This and the fact that the form of $x$ here is the same as the Seiberg-Witten differential
  of the $\CN=2^*$ gauge theory where $\CF_0^{{\mathcal C}_{1,1}}$ is changed to the prepotential 
  (see \cite{FL, MT})
  lead to the conclusion that the free energy in the large $N$ limit of this matrix model is
  exactly the same as the prepotential of the gauge theory (under the identification above).

\section{Conclusions}
\label{sec:conclusion}

In this paper we proposed that the $\beta$-deformation of matrix models provides, in a suitable limit, 
the quantization of the associated integrable system. In particular we have shown that the loop equations
for the $\beta$-deformed generalized matrix models \cite{MY,BMTY} reproduce in the Nekrasov-Shatashvili limit
the Hamiltonians of the quantum Hitchin system associated to the sphere and torus with marked points.
Moreover, we have shown how to obtain the wave function from degenerate field insertions.

It would be interesting to understand if this procedure could provide a general quantization prescription of
 integrable systems which can be linked to specific matrix models.
To this end it would be very useful to provide further evidence and examples. For instance, it would be interesting 
to investigate the $\beta$-deformed Chern-Simons matrix model \cite{Brini:2010fc} in this direction.
Furthermore, the extension of our approach to q-deformed conformal blocks, along the lines of \cite{AY}, would be worth
to be analyzed with the aim of connecting our results with topological strings.

On a more specific side, a natural extension of our analysis concerns Hitchin systems on 
curves of higher genera, the point being a generalization of the identity (\ref{pformula}).
A further explorable direction would be the extension to higher rank gauge groups with
a multi-matrix model approach.

The problem of understanding the proper quantization of the Seiberg-Witten geometry has been explored recently
also from a different view point consisting in a saddle point analysis of the instanton partition in the Nekrasov-Shatashvili
limit \cite{Poghossian:2010pn,Fucito:2011pn,Chen:2011sj}. In Section  \ref{sec:torus} we have seen that 
for the ${\cal N}=2^*$ theory the eigenvalues of the Hamiltonian match. It would be interesting to further explore the relation 
between the two quantizations.

\section*{Acknowledgements}
We thank  
A.~Brini,
B.~Dubrovin,
H.~Itoyama,
S.~Pasquetti,
F.~Yagi,
for interesting discussions and comments.

G.B. and K.M. are partially supported by the INFN project TV12. 
A.T. is partially supported by PRIN ``Geometria delle variet\'a algebriche e loro spazi di moduli'' and the INFN project PI14
``Nonperturbative dynamics of gauge theories''.


\appendix

\section*{Appendix}

\section{Elliptic functions}
\label{sec:}
  The elliptic theta function is defined by
    \bea
    \vartheta_1(z|\tau)
     =     2 q^{1/8} \sin z \prod_{n=1}^\infty (1 - q^n) (1 - 2q^n \cos 2 z + q^{2n})
    \eea
  which has pseudo periodicity
    \bea
    \vartheta_1(z + \pi|\tau)
     =   - \vartheta_1(z|\tau), ~~
    \vartheta_1(z + \pi \tau|\tau)
     =     e^{- i (2 z + \pi \tau)} \vartheta_1 (z|\tau).
    \eea
  This function satisfies $\vartheta_1^{''}(z|\tau) = - 8 \frac{\partial}{\partial \ln q} \vartheta_1(z|\tau)$
  where $\vartheta'_1(z|\tau) = \frac{\partial}{\partial z} \vartheta_1(z|\tau)$.
  
  The Weierstrass elliptic function $\CP$ is double periodic with periods $\pi$ and $\pi \tau$
  and is expressed as
    \bea
    \CP(z)
    &=&  - \zeta'(z), ~~~
    \zeta(z)
     =     \frac{\vartheta'_1(z|\tau)}{\vartheta_1(z|\tau)} + 2 \eta_1 z, ~~~
    \eta_1
     =   - \frac{1}{6} \frac{\vartheta'''_1(z|\tau)|_{z=0}}{\vartheta'_1(z|\tau)|_{z=0}},
           \label{pe}
    \eea
  where $\vartheta_1(z|\tau)$ is elliptic theta function.
  The Weierstrass function satisfies
    \bea
    \CP(z)'
     =     4 \CP(z)^3 - g_2 \CP(z) - g_3,
    \eea
  where 
    \bea
    g_2
    &=&    \frac{4}{3} 
           \left( 1 + 240 \sum_{n=1}^\infty \frac{n^3 q^{n}}{1 - q^{n}} \right),
           \nonumber \\
    g_3
    &=&    \frac{8}{27} 
           \left( 1 - 504 \sum_{n=1}^\infty \frac{n^5 q^{n}}{1- q^{n}} \right),
           \label{g2g3}
    \eea
  We also define $g_1$ whose expansion is 
    \bea
    g_1
     =   - \frac{1}{3} 
           \left( 1 - 24 \sum_{n=1}^\infty \frac{n q^{n}}{1- q^{n}} \right),
           \label{g1}
    \eea
  which is related with $\eta_1$ as $g_1 = - 2 \eta_1$:
    \bea
    \eta_1
     =     \frac{1}{6} \left( 1 - 24 \sum_{n=1}^\infty \frac{n q^{n}}{1- q^{n}} \right)
     =     4 \frac{\partial}{\partial \ln q} \ln \eta,
    \eea
  where $\eta = q^{1/24} \prod_{n=1}^\infty (1 - q^n)$ is the Dedekind eta function.
  
  The Weierstrass and zeta functions satisfies the identity
    \bea
    \CP(a+b) + \CP(a) + \CP(b)
     =     (\zeta(a + b) - \zeta(a) - \zeta(b))^2.
           \label{pformula}
    \eea
  This leads to the following formula of the theta function
    \bea
    \frac{\vartheta_1'(b-a)}{\vartheta_1(b-a)} \left( \frac{\vartheta_1'(a)}{\vartheta_1(a)} - \frac{\vartheta_1'(b)}{\vartheta_1(b)} \right)
     =     \frac{\vartheta_1'(a)}{\vartheta_1(a)}\frac{\vartheta_1'(b)}{\vartheta_1(b)}
         - \frac{1}{2} \left( \frac{\vartheta_1^{''}(a)}{\vartheta_1(a)} + \frac{\vartheta_1''(b)}{\vartheta_1(b)}
         + \frac{\vartheta_1''(b-a)}{\vartheta_1(b-a)} \right) - 3 \eta_1.
           \label{thetaformula}
    \eea


\end{document}